\documentstyle[12pt]{article}
\begin{document}             
\global\arraycolsep=2pt 
%
\thispagestyle{empty} 
\begin{titlepage}    
\topskip 4cm
\begin{center}
{\Large\bf Infrared divergence on the light-front
             and dynamical Higgs mechanism}
\end{center}                                   
            
\vspace{0.8cm}
              
\begin{center}
Shinji Maedan            
   \footnote{  FAX  81-426-68-5094,    TEL  81-426-68-5154,
                     E-mail: maedan@tokyo-ct.ac.jp}   
           \\
\vspace{0.8cm}
{\sl  Department of Physics, Tokyo National College of Technology,
        Kunugida-machi, Hachioji-shi, Tokyo 193, Japan
                                   }
\end{center}                                               
            
\vspace{0.5cm}
              
\begin{abstract}
\noindent       
Dynamical Higgs mechanism on the light-front (LF) is studied using
   a $ (1+1)-$dimensional model, with emphasis on the infrared divergence
problem.
The consideration of the zero mode $ k^+ = 0 $ is not sufficient for
investigating
   dynamical symmetry breaking on the LF.
It also needs to treat properly an infrared divergence
   caused by internal momentum
   $ p^+ \rightarrow 0 ~ ( p^+ \neq 0 ) $ in the continuum limit.
In order to avoid the divergence, we introduce an infrared cutoff function
   $ F_{\rm IR} ( p, \Lambda ) $
   which is not Lorentz invariant.
It is then shown that the gauge boson obtains
   mass dynamically on the LF.
\end{abstract} 
\vskip 2.5cm
\begin{center}
PACS numbers: 11.10.Ef, 11.15.Ex, 11.15.Pg
\vskip 0.1cm
Keywords:  Light-front, Higgs mechanism, Dynamical symmetry breaking,\\
           Infrared divergence, Lorentz invariance
\end{center}
\vfill            
\end{titlepage}
%
%
\baselineskip=0.7cm
%
%
\setcounter{page}{1}
\section{Introduction}

In the light-front (LF) quantized field theory
   \cite{rf:BroPauPin}, LF momentum
   $ k^+ \equiv ( k^0 + k^1 )/ \sqrt 2 $ can take only the
   positive semi-definite value $ k^+ \geq 0 $.
This property brings attractive features to the LF theory,
   one of which is the simplicity of a
   vacuum state.
If particles are all massive, the Fock vacuum
  is the ground state of the system.

The problem of how spontaneous symmetry breaking (SSB) can take place on the
   LF with such a simple vacuum state has
   been discussed [2-14],
   and it has been revealed that the zero mode $( k^+ =0 )$
   is responsible for SSB in many cases.
There are two types of zero mode, one is the constraint zero mode
    \cite{rf:MasYam},
   the other is the dynamical zero mode \cite{rf:KalPauPin}.
In order to treat the zero mode clearly, it is convenient to put
   the system in a "box" with finite size, $  -L \leq x^- \leq L $,
   and finally take the continuum limit $ L \rightarrow \infty $
    \cite{rf:MasYam}.
By solving the zero mode constraint with approximations, it is shown
   that SSB occurs on the LF in simple models.

However, there are few investigations of the Higgs mechanism on the LF.
Therefore, in this paper, we study dynamical Higgs mechanism on the LF using
   a $(1+1)-$dimensional model once proposed by Gross and Neveu
    \cite{rf:GroNev}.
In this model, SSB occurs by quantum effect (fermion's one loop)
   and the gauge boson obtains mass dynamically in the large $N$ limit.
The model has the merit that the zero mode constraint, to leading order
   in $1/N$, contains no operator part, so that we can solve it
   to leading order without worrying about operator ordering.

While the zero mode is important to describe such dynamical SSB on the LF,
   it is not enough.
In addition, we must analyze carefully an infrared divergence
   caused by internal momentum $ p^+ \rightarrow 0 ~ ( p^+ \neq 0 ) $
   in the continuum limit
   $ L \rightarrow \infty $
    \cite{rf:ThiOht,rf:TsuYam,rf:KojSakSak,rf:ItaMae,rf:Ita,rf:NakYam},
   which divergence is peculiar to the LF.
In our model, such infrared divergences $ p^+ \rightarrow 0  $ appear
   in some one-loop fermion's diagrams, even though
   massive fermion.
In order to avoid the divergence, we introduce an infrared
   cutoff function $ F_{\rm IR} ( p, \Lambda ) $ which is not Lorentz
   invariant.
By use of $ F_{\rm IR} $, we show that the gauge boson obtains mass
   dynamically on the LF.

Here we neglect the dynamical zero mode of the gauge field
   and the winding number of a
   complex scalar field.
Instead, without these, we investigate dynamical Higgs mechanism
   by considering the constraint zero modes 
   with careful treatment of the infrared divergence on the LF.
If instantons or $\theta$ vacuum
   existed in a model, the dynamical zero mode would play an important role
   \cite{rf:KalRob,rf:HarOkaTan}.
This problem will be discussed elsewhere in the $(1+1)-$dimensional
   Abelian-Higgs model on the LF \cite{rf:ItaMaeTac}.
%
%
%
%
%
\section{The model and zero mode constraint}
Once Gross and Neveu proposed a $ (1+1)-$dimensional model
\begin{eqnarray}
  {\cal L} &=& ~ \sum_{a=1}^{N} \bar \psi^a~( i \rlap/ \partial~+e \rlap/ B 
                     \gamma_5 )  \psi^a~ 
                - {1 \over 4}( \partial_\mu B_\nu - \partial_\nu B_\mu )
                            ( \partial^\mu B^\nu - \partial^\nu B^\mu ) 
                   \nonumber  \\
          & &   -{N \over 2 \lambda} (\sigma^2 +\pi^2)
              - \sum_a \bar \psi^a~( \sigma + i \gamma_5 \pi )  \psi^a ,
              \hskip 1cm  ( \mu = 0,~1 )
  \label{ba}
\end{eqnarray}
   which exhibits dynamical Higgs mechanism in the large $ N $ limit
    \cite{rf:GroNev}.
$ \psi^a $ is an $ N- $component massless fermion, $ B_\mu $ is a $ U(1) $
   gauge field, $ \sigma $ is a scalar and $ \pi $ is a pseudoscalar field.
A bare coupling constant $\lambda$ is of order $ O(N^0) $,
   and $e$ is $ O(1 / \sqrt N) $.
This lagrangian (\ref{ba}) is invariant under
  the local chiral gauge transformations,
\begin{eqnarray}
  \psi^a  & \rightarrow &  {\rm exp} \left\{ i \gamma_5 \theta (x)
                           \right\} \psi^a,   \nonumber  \\
  B_\mu   & \rightarrow &  B_\mu + {1 \over e }~ \partial_\mu \theta (x),
                                    \nonumber  \\
   \left(
  \begin{array}{c}
     \sigma  \\
     \pi
   \end{array}
   \right)
   &   \rightarrow   &
   \left(
   \begin{array}{cc}
      {\rm cos}~ 2 \theta (x)   &  {\rm sin}~ 2 \theta  (x) \\
      -{\rm sin}~ 2 \theta (x)  &  {\rm cos}~ 2 \theta  (x)
   \end{array}
   \right)
   \left(
   \begin{array}{c}
     \sigma  \\
     \pi
   \end{array}
   \right).
  \label{bb}
\end{eqnarray}

Before discussing the LF formalism, we will see briefly how the gauge boson
   $ B_\mu $
   obtains mass dynamically in the equal-time quantization formalism.
Our interest is now the condensation of $ \sigma $ or $ \pi $ , so we 
   integrate (\ref{ba}) over the fermion field $ \psi^a $
\begin{eqnarray}
  {\cal L} &=& - {1 \over 4}( \partial_\mu B_\nu - \partial_\nu B_\mu )
                           ( \partial^\mu B^\nu - \partial^\nu B^\mu ) 
               -{N \over 2 \lambda} ( \sigma^2 +\pi^2 ) \nonumber  \\
     & & \hskip 1cm   - ~i N ~ {\rm Tr} ~ {\rm log}
                    ~( i \rlap/ \partial~-\sigma - i \gamma_5 \pi +e \rlap/ B 
                      \gamma_5 ) ~  .
 \label{bc}
\end{eqnarray}
To leading order in $ 1/N $, the effective potential of
   $ \sigma $ and $ \pi $ becomes double well type due to fermion one loop
   correction.
Hence $ \sigma $ field has a vacuum expectation value $ v $, and $ \pi $ field
   is now a
   would-be-Goldstone boson.
The massless gauge field $ B_\mu $ combines with this $ \pi $ field to
   become a massive vector field with mass $ e {\sqrt N} / \sqrt \pi $~
    \cite{rf:GroNev}.

Now, within the framework of the LF formalism, we study the lagrangian 
   (\ref{bc}) in which the fermion field has been integrated.
This is because the condensation of $\sigma$ or $\pi$ is of interest
   when investigating the Higgs mechanism in our model.
If the system is put in a "box" ( i.e. $ - L \leq x^- \leq L $ ),
   momentum $k^+$ is discretized and one can isolate the zero mode clearly
    \cite{rf:MasYam}.
We use such a discretization method for the simple reason that the
   zero mode should be separated from other modes in order to
   examine SSB on the LF.
Therefore, it should be taken into account that the continuum limit
   $ L \rightarrow \infty $ is taken after calculations are over.
Boundary conditions of $ \sigma (x), \pi (x) $ and $ B_\mu (x) $
 are chosen to be periodic.

Let us parametrize the fields $ ( \sigma, \pi ) $
   in polar variables $ ( \xi, \eta ) $ such that
   $ \sigma + i \pi = \xi~ e^{i \eta} $, and furthermore rescale the gauge
   field $ B_\mu $ and the coupling constant $ e $ as 
   $ B_\mu \rightarrow \sqrt{N} B_\mu $ and
   $ e \rightarrow  e / \sqrt{N} $.
Consequently, the lagrangian is rewritten as
\begin{eqnarray}
  {\cal L} &=& - {N \over 4}( \partial_\mu B_\nu - \partial_\nu B_\mu )
                           ( \partial^\mu B^\nu - \partial^\nu B^\mu ) 
               -{N \over 2 \lambda }~ \xi^2        \nonumber  \\
           & & \hskip 1cm     - ~i N ~ {\rm Tr} ~ {\rm log}
                  ~( i \rlap/ \partial~ - \xi ~ {\rm exp} (  i \gamma_5 \eta )
                        +e \rlap/ B \gamma_5 )  ~,
 \label{bca}
\end{eqnarray}
where $ e \sim O( N^0 ) $.

Hereafter, we shall derive the zero mode constraints for $\xi (x)$
   and $\eta (x)$ to leading order in $ 1/ \sqrt N$.
As mentioned in Section 1, we neglect the dynamical zero mode of the gauge
   field $ B_\mu $ and the winding number $ 2 \pi m = \eta (L) - \eta (-L) $
   of the phase field.
Hence, the polar variables $\xi (x)$ and $\eta (x)$ are also taken to satisfy
   periodic boundary conditions.
Separate c-number parts  $ \xi_0 $ and $\eta_0 $ of the zero mode $( k^+ =0 )$
   of  $ \xi (x) $ and $\eta (x) $, respectively,
\begin{eqnarray}
   \xi (x)  = \xi_0 + \tilde \xi (x),  \nonumber    \\
   \eta (x) = \eta_0 + \tilde \eta (x) ~.
  \label{bf}
\end{eqnarray}
Note that  $ \tilde \xi (x) $ and $ \tilde \eta (x) $ contain both
   operator valued zero mode and oscillation modes.
With the equation (\ref{bf}), the Euler-Lagrange equation for $\xi$ is
   expressed as \cite{rf:BorGraWer,rf:Mae}
\begin{eqnarray}
   0 &=&   {N \over \lambda}~ \xi - i N~ {\rm Tr}~
              ~ {\rm exp} (  i \gamma_5 \eta )~ 
               \left\{~ i \rlap/ \partial~ - \xi ~ {\rm exp} ( i \gamma_5
\eta ) 
                  +e \rlap/ B  \gamma_5 ~\right\}^{-1}  \nonumber  \\
    &=& {N \over \lambda}~ (\xi_0 + \tilde \xi ~)
         - i N~ {\rm Tr}~ {\rm exp} (  i \gamma_5 (\eta_0 + \tilde \eta ))~ 
                                   \nonumber   \\
    & & \hskip 0.1cm \times \biggl\{~S_0 +S_0 \left[ {\rm exp} ( i \gamma_5
\eta_0 )
            \{ \xi_0 ( {\rm exp} ( i \gamma_5 \tilde \eta ) -1 )
               + \tilde \xi {\rm exp} ( i \gamma_5 \tilde \eta ) \}
           - e \rlap/ B  \gamma_5 \right] S_0  \biggr.   \nonumber   \\
    & & \hskip 1.1cm  \biggl. + S_0 ~ [ \cdots ] ~ S_0 ~ [ \cdots ] ~ S_0
                             + \cdots ~ \biggr\} ~,
  \label{bd}
\end{eqnarray}
where $ S_0 (x,y) $ is defined as
\begin{equation}
  S_0 (x,y) \equiv (i {\rlap/ \partial} 
             - \xi_0 ~ {\rm exp} (i \gamma_5 \eta_0 ) + i \epsilon )^{-1}~.
  \label{bk}
\end{equation}

After substituting (\ref{bf}) into the lagrangian (\ref{bca}), one can
   easily find that the $ \tilde \xi (x) $ propagator is proportional
   to $ 1/N $.
Accordingly, the order of $ \tilde \xi (x) $ is $ O(1/ \sqrt N) $ at most, and
   we expand the field in terms of $ 1/ \sqrt N $ as
    \cite{rf:BorGraWer,rf:Mae}
\begin{equation}
   \tilde \xi (x)  = \tilde \xi^{(1)} (x) + \tilde \xi^{(2)} (x) + \cdots~,
  \label{bg}
\end{equation}
where the order of $ \tilde \xi^{(1)} (x) $ is $ 1/ \sqrt N $, 
   the order of $ \tilde \xi^{(2)} (x) $ is $ 1/ N $, and so on.
For the same reason, $ \tilde \eta (x) $ and $ B_\mu (x) $ are expanded
   in terms of $ 1/ \sqrt N $ as
\begin{eqnarray}
   \tilde \eta (x)  = \tilde \eta^{(1)} (x) + \tilde \eta^{(2)} (x) + \cdots~,
                                      \nonumber  \\
   B_\mu (x) = \tilde B_\mu ^{(1)} (x) +  \tilde B_\mu ^{(2)} (x) + \cdots~.
  \label{bh}
\end{eqnarray}
By use of (\ref{bg}) and (\ref{bh}), the Euler-Lagrange equation for $\xi$
   expanded in terms of $ 1/ \sqrt N $ is obtained \cite{rf:BorGraWer,rf:Mae}.

Zero mode constraint for $ \xi $ is derived from the
   Euler-Lagrange equation  (\ref{bd}) by integrating over $ x^- $
    \cite{rf:Rob}.
To leading order in $ 1/ \sqrt N $, this becomes  \cite{rf:BorGraWer,rf:Mae} 
\begin{eqnarray}
  && \int_{-L}^{L} d x^- \left[ N { \xi_0 \over \lambda } - i N~ {\rm tr} ~
          {\rm exp} (i \gamma_5 \eta_0 ) ~ S_0 (x,x) \right] \nonumber  \\
  &&  = (2 L) N  \left[ { \xi_0 \over \lambda } - i ~ {\rm tr} ~
          {\rm exp} (i \gamma_5 \eta_0 ) ~ S_0 (x,x) \right] = 0 ~.
  \label{bi}
\end{eqnarray}
In the same manner, to leading order, zero mode constraint for $\eta$ is
\begin{equation}
  {\rm tr} ~i~ \gamma_5 ~\xi_0 ~ {\rm exp} (i \gamma_5 \eta_0 )~ S_0 (x,x)
                 = 0 ~.
  \label{bj}
\end{equation}
%
%
%
\section{Infrared divergence on the light-front}

Since $ S_0$ , (\ref{bk}), can be regarded as a fermion's propagator,
   momentum in $ S_0 $ is discretized such as
   $ p_n^+ = \pi n / L ~ (~ n = \pm 1/2, \pm 3/2, \cdots ~) $
   where there is no zero mode.
This corresponds to the choice of the antiperiodic boundary condition
   for the fermion field.
The zero mode constraint for $ \xi $ (\ref{bi}) is then given by
\begin{equation}
  { \xi_0 \over \lambda } = 2~ i~ \xi_0 \left( {1 \over 2 L } 
         \sum_{n= \pm \frac{1}{2}, \pm \frac{3}{2}, \cdots ~} \right)
         \int_{- \infty} ^{\infty} { d p^- \over 2 \pi }
              {1 \over 2 p^+ _n p^- - \xi_0 ^2 + i \epsilon } ~,
  \label{ca}
\end{equation}
where internal momentum $ p_n^+ $ does not have zero mode.
The r.h.s corresponds to the one loop diagram of the fermion with
   mass $ \xi_0 $ in the equal-time quantization,
   but we should compute it with the LF metric.

If we naively calculate the $ p^- $ integration in (\ref{ca}) using a 
   Lorentz invariant cutoff function such as
   $ \theta (\Lambda^2 - \vert 2 p_n^+ p^- \vert ) $,
   information of mass $\xi_0$ is lost.
Furthermore, after the continuum limit $ L \rightarrow \infty $, there
   arises an infrared divergence $ p^+ \rightarrow 0 ( p^+ \neq 0 )$
    \cite{rf:ItaMae},
\begin{eqnarray}
   & & {1 \over 2 L } \sum_{n= - L \Lambda / \pi} ^{ L \Lambda / \pi} 
      \int_{- \infty} ^{\infty} { d p^- \over 2 \pi }
        {1 \over { 2 p^+_n \left( p^- - { \xi_0 ^2 \over 2 p_n^+} 
                   + i {\epsilon \over 2 p_n^+} \right)  } } ~
         \theta (~ \Lambda^2 - \vert 2 p_n^+ p^- \vert ~)  \nonumber \\
   & & = - {i \over 2 L } \sum_{n= 1/2} ^{ L \Lambda / \pi}
              {1 \over p_n^+}  ~~ \longrightarrow ~~
              - {i \over 2 \pi} \int_{0}^{\Lambda} { d p^+ \over p^+ }
              \hskip 1cm  ( L \rightarrow \infty ) ~,
  \label{cb}
\end{eqnarray}
where a regularization of an ultraviolet divergence $ p_n^+  \leq \Lambda $
   has been done.
These unfavorable points (i) a loss of mass information, and (ii) the
   infrared divergence $ p^+ \rightarrow 0 ~( p^+ \neq 0 )$, are already
   known as fundamental problems in
   the continuum theory of the LF formalism \cite{rf:NakYam,rf:TsuYam}.
We will comment here that these are common problems on the LF
   and have nothing to do with SSB itself.
   
The renormalization of the infrared divergence was studied by
   Thies and Ohta \cite{rf:ThiOht} in the chirally invariant Gross-Neveu
   model on the LF with no gauge field ($ e=0 $).
They derive the self-consistency condition (the Hartree equation)
\begin{equation}
  1={N g^2 \over 2 \pi} \int_{\epsilon}^{\Lambda} {d p^+ \over p^+},
  \label{cto}
\end{equation}
which is identical to our constraint equation for $\xi$ obtained
   using the cutoff function
   $ \theta (~ \Lambda^2 - \vert 2 p_n^+ p^- \vert ~) $ as  (\ref{cb}).
(Our coupling constant $\lambda$ is related to the coupling constant $g$ in
   Ref \cite{rf:ThiOht} by $ \lambda = N g^2 /2 $.)
The integral in the Hartree equation  (\ref{cto}) is regularized
   by ultraviolet
   and infrared cutoffs.
The ultraviolet divergence is renormalized to an effective
   coupling constant $g_{\rm eff}$.
The Hartree equation (\ref{cto}) is regarded as the infrared
   renormalization condition for the effective
   coupling constant $g_{\rm eff}$.
It should be noted that the ultraviolet cutoff $\Lambda$ and infrared cutoff
   $\epsilon$
   are introduced independently.
Thies and Ohta show that physical quantities do not
   depend on these cutoffs by use of the infrared
   renormalization condition (the Hartree equation).

If one, however, takes notice of the first problem (i) the loss of
   mass information, another management of the infrared divergence
   will be possible.
Namely, the constraint (\ref{ca}) can be regarded as the gap
   equation having a nontrivial solution,
   while it is regarded as the infrared renormalization condition
   in the work by Thies and Ohta.
We now consider the infrared cutoff function by which one can
   interpret (\ref{ca}) as the gap equation.
As discussed previously, the Lorentz invariant cutoff function
   $ \theta ( \Lambda^2 - \vert 2 p_n^+ p^- \vert )$ can not
   prevent the infrared divergence.
At the pole $ p^- = \xi_0^2 / 2 p^+ $, any
   Lorentz invariant cutoff function
   $ G( 2 p^+ p^-, \Lambda ) $ can not depend on $p^+$ because
   $ G( 2 p^+ \times ( \xi_0^2 / 2 p^+), \Lambda ) =
     G( \xi_0^2, \Lambda ) $.
Thereby, $ G( 2 p^+ p^-, \Lambda ) $ does not restrict the region of
   internal momentum $p^+$ in the Feynman integral, and there arises the
   infrared divergence as seen from (\ref{cb}).
As long as a Lorentz invariant regularization is used, we can not
   overcome these difficulties.
In order to avoid these, it is almost inevitable to violate the
   Lorentz invariance \cite{rf:NakYam}.
   
Therefore, we introduce a Lorentz noninvariant infrared
   cutoff function $ F_{\rm IR} ( p,\Lambda ) $ such as
\begin{equation}
  F_{\rm IR} ( p,\Lambda ) =
    \left\{ - \Lambda^2 \over (p^-)^2 - \Lambda^2 + i \epsilon \right\}~.
  \label{cc}
\end{equation}
Since this $ F_{\rm IR} $ cuts off $ p^- $ at $\Lambda$, it effectively
   brings an infrared cutoff $ p^+ \sim \xi_0^2 / 2 \Lambda $ on
   mass shell $ 2 p^+ p^- = \xi_0^2 $.
It is similar to the parity invariant regularization
    \cite{rf:Ita,rf:HarVar}  where the ultraviolet and infrared cutoffs are
    related to each other.
Note that the effective infrared cutoff of $p^+$ depends on the mass $\xi_0$,
   but $ F_{\rm IR} $ (\ref{cc}) does not involve $\xi_0$.
   
We shall regularize the ultraviolet divergence by the replacement
  $
  ( p^2 - \xi_0^2 + i \epsilon )^{-1} ~ \rightarrow ~
  ( p^2 - \xi_0^2 + i \epsilon )^{-1} -
  ( p^2 - \Lambda^2 + i \epsilon )^{-1} 
  $,
then a properly regularized expression of the zero mode constraint
   (\ref{ca}) is
\begin{eqnarray}
  { \xi_0 \over \lambda }
   &=&  2~ i~ \xi_0 \left( {1 \over 2 L } 
         \sum_{n= \pm \frac{1}{2}, \pm \frac{3}{2}, \cdots ~}  \right)
         \int_{- \infty} ^{\infty} { d p^- \over 2 \pi }
           \left( {1 \over 2 p^+ _n p^- - \xi_0 ^2 + i \epsilon }
                 -{1 \over 2 p^+ _n p^- - \Lambda ^2 + i \epsilon } \right)
                                 \nonumber   \\
   & &  \hskip 5.5cm  \times \left\{ - \Lambda^2 \over
                      (p^-)^2 - \Lambda^2 + i \epsilon \right\} \nonumber  \\
   &=&  2~ i~ \xi_0 \left( {1 \over 2 L }
        \sum_{n= \frac{1}{2}, \frac{3}{2},\cdots ~}  \right)
           \left( {1 \over 2 p^+_n + {\xi_0 ^2 \over \Lambda} }
                 -{1 \over 2 p^+_n + {\Lambda^2 \over \Lambda} } \right) (-i)
                                   \nonumber  \\
   & \longrightarrow & 
       2 i \xi_0 
         \int_{0} ^{\infty} { d p^+ \over 2 \pi }
           \left( {1 \over 2 p^+ + {\xi_0 ^2 \over \Lambda} }
                 -{1 \over 2 p^+ + {\Lambda^2 \over \Lambda} } \right) ( -i )
       ~=~ {\xi_0 \over 2 \pi} {\rm log}{\Lambda^2 \over \xi_0^2} ~,
  \label{ce}
\end{eqnarray}
where we have taken the continuum limit $ L \rightarrow \infty $
   because of the infrared cutoff function $ F_{\rm IR} $.
Physical quantities such as a vacuum expectation value $\xi_0$
   should not depend on the artificial parameter $L$; indeed,
   $\xi_0$ has no $L$ dependence in the continuum limit.
   
This zero mode constraint (\ref{ce}) in the continuum limit is nothing but
   the gap equation \cite{rf:BorGraWer,rf:Mae}.
It has a nontrivial solution
\begin{equation}
  \xi_0 = M {\rm exp} (1-\pi / \lambda_R )  > 0~, \hskip 1cm
  {1 \over \lambda_{\rm R} } \equiv {1 \over \lambda }
     + \frac{ 1 }{2 \pi} {\rm log} {M^2 \over \Lambda^2} + \frac{ 1 }{ \pi}~,
  \label{cf}
\end{equation}
where $ \lambda_{\rm R} $ is a renormalized coupling constant and 
   $M$ is a renormalization point \cite{rf:GroNev,rf:ItaMae}.
(Our notation $ \xi_0 $ corresponds to $ M_{\rm F} $ in
   Ref \cite{rf:GroNev}.)
Although a trivial solution $ \xi_0 =0 $ also exists, it is not suitable
   because of a tachyon pole of the $ \tilde \xi (x) $ propagator.
   
On the other hand, the $ \eta $'s zero mode constraint (\ref{bj}) whose
   l.h.s is always zero allows  $ \eta_0 $ to have any real value.
This reflects the fact that all the points on a circle with radius
   $\xi_0$ in the $( \sigma, \pi )$ plane are
   physically equivalent due to the chiral invariance of the model.
We choose a solution
\begin{equation}
  \eta_0 = 0,
  \label{cg}
\end{equation}
of the zero mode constraint for simplicity.

In the equal-time quantization formalism, the chiral equivalent vacua
   are degenerate and one chooses a vacuum state among them.
On the other hand, in the LF formalism, a vacuum state is the Fock vacuum
   and one chooses a solution of the zero mode constraint.
Multi-vacuum states in the equal-time formalism correspond to multi-solutions
   of the
   zero mode constraint in the LF formalism
    \cite{rf:Rob,rf:PinSanHil}.
%
%
%
%
%
%
\section{Dynamical Higgs mechanism}

To leading order in $ 1/ \sqrt N $,  the effective lagrangian
   involving the gauge field $ \tilde B_\mu (x) $ and the phase
   $ \tilde \eta (x) $ is obtained from the lagrangian (\ref{bca})
   with the help of (\ref{bf}), (\ref{cf}) and  (\ref{cg}) 
\begin{eqnarray}
  {\cal L}_{\rm eff} ( \tilde B_\mu , \tilde \eta )
   & = & -{N \over 4} 
           ( \partial_\mu \tilde B_\nu^{(1)} - \partial_\nu \tilde
B_\mu^{(1)} )^2
                                          \nonumber  \\
   & & \hskip 0.1cm +i N \left[ -{1\over 2} ~\xi_0 ~ {\rm Tr} ~ S ~
          \tilde \eta^{(1)} \tilde \eta^{(1)}  ~
          + {1 \over 2} ~ \xi_0^2 ~ {\rm Tr}
                 (~ S ~ i ~ \gamma_5 ~ \tilde \eta^{(1)} ~ S ~ i ~ \gamma_5
                         \tilde \eta^{(1)} )  \right.  \nonumber  \\
  & &  \hskip 0.2cm -  {1 \over 2}~ e ~ \xi_0 {\rm Tr} (~S ~ i ~ \gamma_5 ~
                    \tilde \eta^{(1)} ~ S ~ {\rlap/ \tilde B}^{(1)}~ \gamma_5 ~)
         -  {1 \over 2} e ~ \xi_0 {\rm Tr} (~ S {\rlap/ \tilde B}^{(1)} ~
\gamma_5
                        ~ S ~ i ~ \gamma_5 ~ \tilde \eta^{(1)} ~)   
\nonumber  \\
   & & \hskip 0.2cm  \left. +  {1 \over 2} e^2 ~ {\rm Tr}
           (~ S ~ {\rlap/ \tilde B}^{(1)} ~
                      \gamma_5 ~ S ~ {\rlap/ \tilde B}^{(1)} ~ \gamma_5 ~) 
                                \right]     
         +( \rm constant~~ term )~,          \nonumber  \\
   & & \hskip 7cm  ( \mu, \nu = +, - ) 
  \label{da}
\end{eqnarray}
where $ S $ is defined as
\begin{equation}
  S (x,y) \equiv \left.  S_0 (x,y) \right|_{ \eta_0 = 0 } ~
  = (i {\rlap/ \partial} 
             - \xi_0 ~ + i \epsilon )^{-1}~.
  \label{db}
\end{equation}
To derive $ {\cal L}_{\rm eff} $, the $ 1/ \sqrt N $ expansion  (\ref{bg}) 
   and  (\ref{bh}) also have been used.
One loop Feynman integrals on the LF in  (\ref{da}) are calculated as follows.

$ {\rm Tr} S ( x,x ) $ having both ultraviolet and infrared divergence
   is similar to the r.h.s of the gap equation  (\ref{ce}).
It then becomes
   $ (-i / 2 \pi)~\xi_0 ~  {\rm log}{\Lambda^2 / \xi_0^2} ~$
   in the continuum limit.
Next, $ {\rm Tr} (~ S ~ i ~ \gamma_5 ~ \tilde \eta^{(1)} ~ S ~ i ~ \gamma_5
   \tilde \eta^{(1)} ) $
   has no infrared divergence but it diverges in the ultraviolet region.
We regularize it by the replacement
   $ ( \rlap/ p - \xi_0 + i \epsilon )^{-1} \rightarrow
     ( \rlap/ p - \xi_0 + i \epsilon )^{-1} 
    - ( \rlap/ p - \Lambda + i \epsilon )^{-1} $,
   so
\begin{eqnarray}
  & &  {\rm tr} \left( {1 \over 2 L } \sum_{n= \pm \frac{1}{2}, \cdots } \right)
         \int_{- \infty} ^{\infty} { d p^- \over 2 \pi } 
           \left( {1 \over {\rlap/ p} - \xi_0 + i \epsilon }
                 - {1 \over {\rlap/ p} - \Lambda + i \epsilon }\right)
                    i \gamma_5               \nonumber   \\
   & &  \hskip 1cm    \times 
           \left( {1 \over ( {\rlap/ p}-{\rlap/ k} ) - \xi_0 + i \epsilon }
                 - {1 \over ( {\rlap/ p}-{\rlap/ k} )
                      - \Lambda + i \epsilon} \right)  i \gamma_5  ~,
  \label{dd}
\end{eqnarray}
with discretized internal LF momentum $ p_n^+ = \pi n / L $ and external LF
   momentum $ k^+ > 0 $.
Positions of  $p^-$ poles of the integrand in  (\ref{dd}) depend on
   the value of momentum $p_n^+$ \cite{rf:LigBak}.
(i)  $ p_n^+ < 0 $ ; all poles are in the upper half-plane,
(ii) $ 0 < p_n^+ < k^+ $ ; two poles are in the upper half-plane and the other
   two poles are in the lower half-plane,
(iii) $ p_n^+ > k^+ $ ; all poles are in the lower half-plane.
Then, in the continuum limit,  (\ref{dd}) becomes
\begin{eqnarray}
  && \lim_{L \rightarrow \infty}
     \left( {1 \over 2 L } \sum_{n= \pm \frac{1}{2}, \cdots} \right) 
           ~\theta~(p_n^+) ~\theta~(k^+ - p_n^+)
           \int_{- \infty} ^{\infty} { d p^- \over 2 \pi }
            2 (\xi_0-\Lambda)^2      \nonumber \\
  & & \hskip 0.2cm  \times \left[
      { 
        {
           (\xi_0+\Lambda)^2 (p^2- p \cdot k)-(p^2+\xi_0 \Lambda)
                                 \{ (p-k)^2 + \xi_0 \Lambda \} 
                                                                 }
        \over
        {  
           (p^2-\xi_0^2+i \epsilon)(p^2-\Lambda^2+i \epsilon)
             \{ (p-k)^2 - \xi_0^2+i \epsilon \}
             \{ (p-k)^2 - \Lambda^2+i \epsilon \}
                                                                 }
                                                                      }
                                      \right]     \nonumber   \\
   &=& -{i \over 2 \pi}~ {\rm log}~{\Lambda^2 \over \xi_0^2}
       - {{i \over \pi} {\sqrt {k^2 \over 4 \xi_0^2 - k^2} }~
             {\rm arctan} {\sqrt {k^2 \over 4 \xi_0^2 - k^2} }}
       + O \left( { {\rm log} \Lambda \over \Lambda } \right) ~.
  \label{de}
\end{eqnarray}
The first term which diverges ultravioletly cancels with $ {\rm Tr} ~S $
   in $ {\cal L}_{\rm eff} $.
The term  \newline
   $ {\rm Tr} (~S ~ i ~ \gamma_5 ~  \tilde \eta^{(1)} ~ S ~
     {\rlap/ \tilde B}^{(1)}~ \gamma_5 ~) $
   suffers from no divergence and can be calculated with no problem.
   
Since the vacuum polarization term
   $  {\rm Tr} (~ S ~ {\rlap/ \tilde B}^{(1)} ~
     \gamma_5 ~ S ~ {\rlap/ \tilde B}^{(1)} ~ \gamma_5 ~) $
   appears to diverge ultraviolatly, we take the regularization
   $ ( \rlap/ p - \xi_0 + i \epsilon )^{-1} \rightarrow
     ( \rlap/ p - \xi_0 + i \epsilon )^{-1} 
    - ( \rlap/ p - \Lambda + i \epsilon )^{-1} $
   as in  (\ref{dd}).
Among four components, only $ (-,-) $ component
   $  {\rm Tr}~  S ~( \tilde B_{-}^{(1)} ~\gamma^{-} )~
         \gamma_5 ~ S ~( \tilde B_{-}^{(1)} ~\gamma^{-} ) ~ \gamma_5 ~ $
   contains  infrared divergent terms in the continuum limit,
\begin{eqnarray}
  & &  2 \left( {1 \over 2 L } \sum_{n= \pm \frac{1}{2}, \cdots} \right)
         \int_{- \infty} ^{\infty} { d p^- \over 2 \pi } 
         ~( 2 p^- p^- )
           \left( {1 \over p^2 - \xi_0^2 + i \epsilon }
                 - {1 \over p^2 - \Lambda^2 + i \epsilon }\right)
                                            \nonumber   \\
   & &  \hskip 2cm    \times 
           \left( {1 \over ( p - k )^2 - \xi_0^2 + i \epsilon }
                 -{1 \over ( p - k )^2 - \Lambda^2 + i \epsilon} \right)~.
  \label{dfa}
\end{eqnarray}
After $p^-$ integration in (\ref{dfa}), one can find there are infrared
divergent
   terms as $ L \rightarrow \infty $.
This divergence is the peculiar character of the LF formalism,
   which is already observed in the gap equation  (\ref{cb}).
In order to avoid it, we make use of the infrared
   cutoff function $ F_{\rm IR} $ (\ref{cc}) as in the gap equation,
   then the term (\ref{dfa}) with $ F_{\rm IR} $ becomes
\begin{eqnarray}
  & & \lim_{L \rightarrow \infty} \left\{
       {\rm Eq.( \ref{dfa} )} \times F_{\rm IR} ( p, \Lambda ) \right\}
                                        \nonumber   \\
  &=&  {i \over \pi}~{k^- k^- \over k^2}
       + {2i \over \pi}~{k^- k^- \over k^2}
          { ( k^2 - 2 \xi_0^2 ) \over
              \sqrt {k^2 ( 4 \xi_0^2 - k^2 ) }  }
             {\rm arctan} {\sqrt {k^2 \over 4 \xi_0^2 - k^2} }
       + O \left( { {\rm log} \Lambda \over \Lambda } \right) ~.     
  \label{dfb}
\end{eqnarray}

Consequently, in the continuum limit $ L \rightarrow \infty $
   and $ \Lambda \rightarrow \infty $, the
   effective lagrangian  (\ref{da}) in momentum space results
\begin{eqnarray}
  &&{\cal L}_{\rm eff} / N
   = {1 \over 2} \tilde B_{\mu}^{(1)}  \left\{
       ( k^2 - {e^2 \over \pi} ) 
          \left( -g^{\mu \nu}+{ k^\mu k^\nu \over k^2} \right) \right\}
                   \tilde B_{\nu}^{(1)}  \nonumber  \\
  &&  \hskip 1cm - e^2 ~ \xi_0^2 ~{ U(k^2) \over k^2 }~( k^\mu )
       \left\{  \tilde B_{\mu}^{(1)} - {i \over 2 e} ( k_\mu ) \tilde \eta^{(1)}
                                            \right\}      (-k^\nu)
       \left\{  \tilde B_{\nu}^{(1)} - {i \over 2 e} (-k_\nu) \tilde \eta^{(1)}
                                            \right\} ~,
  \label{dh}
\end{eqnarray}
with a function
   $ U(k^2) \equiv
       2~ \left\{ \pi \sqrt {k^2 ( 4 \xi_0^2 - k^2) }~ \right\}^{-1}
        {\rm arctan} \sqrt { k^2 / (4 \xi_0^2 - k^2) }      $~.

We choose the unitary gauge
\begin{equation}
   \tilde A_{\mu}^{(1)} \equiv  \tilde B_{\mu}^{(1)} 
       + {1 \over 2 e} \partial_\mu  \tilde \eta^{(1)} ~,
  \label{di}
\end{equation}
which is useful to find what are physical particles.
This gives the lagrangian
\begin{equation}
  {\cal L}_{\rm eff} / N
   = {1 \over 2} \tilde A_{\mu}^{(1)}  \left\{
       ( k^2 - {e^2 \over \pi} ) 
          \left( -g^{\mu \nu}+{ k^\mu k^\nu \over k^2} \right)
       + 2 e^2 \xi_0^2 U(k^2) {k^\mu k^\nu \over k^2 }     \right\}
                   \tilde A_{\nu}^{(1)}  ~.
  \label{dj}
\end{equation}
The phase field $ \tilde \eta^{(1)} $ disappears from ${\cal L}_{\rm eff}$.
The equation of motion for $ \tilde A_{\mu}^{(1)} $ 
   is derived from  (\ref{dj}),
\begin{equation}
  ( \partial^\mu \partial_\mu + {e^2 \over \pi } )
              \tilde A_{\nu}^{(1)} = 0 ~,  \hskip 1.5cm
   \partial^\mu  \tilde A_{\mu}^{(1)} = 0 ~,
  \label{dk}
\end{equation}
where $ U(k^2) \neq 0 $ has been used.

Thus we have shown that the gauge boson obtains mass $ e/ \sqrt \pi $
   (with the original $e$, $ e {\sqrt N} / \sqrt \pi $ ) dynamically on the LF.
%
%
%
%
\section{Summary}

To summarize, we studied dynamical Higgs mechanism on the light-front.
The complexity of the vacuum is carried by the constraint zero mode.
Zero mode constraint for the (pseudo) scalar field is solved by use of
   the $ 1/N$ expansion.
Its nontrivial solution breaks the symmetry spontaneously.
The choice of a vacuum state among degenerated vacua in the
   equal-time formalism corresponds to the choice
   of a solution of the zero mode constraint
   in the LF formalism.
With the unitary gauge, it is shown that the gauge field obtains mass
   dynamically on the LF in the large $N$ limit.

In calculating the massive fermion's one loop integral, we face the
   infrared divergence $ p^+ \rightarrow 0 ~ ( p^+ \neq 0 ) $ 
   in the continuum limit $ L \rightarrow \infty $.
In order to avoid this infrared divergence,
   which is the peculiar character on the LF, we have introduced
   the Lorentz noninvariant cutoff function
   $ F_{\rm IR} ( p,\Lambda ) =
     - \Lambda^2 / \{ (p^-)^2 - \Lambda^2 + i \epsilon \} $.
On the mass shell, this $  F_{\rm IR} $ effectively brings the
   infrared cutoff $ p^+ \sim \xi_0^2 / 2 \Lambda $
   depending on the mass $\xi_0$~.
In the limit $  L \rightarrow \infty $ and
   $ \Lambda \rightarrow \infty $, the regularized theory recovers
   the Lorentz invariance as seen from (\ref{dh}).

It is essential that, if a cutoff function is Lorentz invariant,
   it is difficult to resolve the problems of the loss of mass information
   and the infrared divergence on the LF.
Careful treatment of the infrared problem is necessary for investigating
   dynamical Higgs mechanism in our model.
\vskip 1.5 cm
\centerline{\bf Acknowledgements}\par
The author would like to thank K. Itakura for valuable discussions.
He also acknowledges stimulating discussions with
   the members of the High Energy Theory Group at Tokyo Metropolitan
   University and Doyo-kai, and M. Tachibana.
%
%
%
\newpage
\end{document}